# Electrical transport properties of high quality Bi-2223 crystal.

Yuri Eltsev[1*], Sergey Lee[2], Koichi Nakao[2], and Setsuko Tajima[2+]

[1]P. N. Lebedev Physical Institute, RAS, Moscow, 119991, Russia

[2]Superconductivity Research Laboratory, ISTEC, 10-13, Shinonome 1-chome, Koto-ku, Tokyo, 135-0062, Japan

Present address: [+]Department of Physics, Osaka University, Toyonaka, Osaka 560-0043, Japan

We have studied electrical transport properties of a high quality $Bi_2Sr_2Ca_2Cu_3O_{10+x}$ crystal below a superconducting critical temperature, $T_c$. In magnetic fields B parallel to the c-axis just above a voltage response onset resistance vs temperature data are well fitted to the vortex-glass model. Obtained from the vortex-glass analysis a melting transition boundary for Bi-2223 crystal shifted towards lower temperatures compared to previously reported data for a (Bi,Pb)-2223 single crystal. The critical current density, $J_c$, of Bi-2223 crystals is close to presented elsewhere $J_c$ values for Bi-2223 tapes suggesting a principal role of weak intrinsic pinning properties of Bi-2223 as a main limiting factor of $J_c$ of Bi-2223 conductors.

---

[*] e-mail: eltsev@sci.lebedev.ru

$Bi_2Sr_2Ca_2Cu_3O_{10+x}$ (Bi-2223) compound with the superconducting critical temperature $T_c \approx 110K$ is one of the promising candidates for practical applications. In recent years substantial progress was achieved in production of long Bi-2223/Ag wires and tapes. But, in spite of great efforts in non-zero magnetic field the critical current density ($J_c$) of these Bi-2223 conductors is rather low at elevated temperatures above 30-40K compared to e.g. YBCO coated conductors [1]. To understand the origin of limited critical current ability of Bi-2223 tapes and wires it is essential to elucidate an effect of extrinsic factors like insufficient densification, poor grain connectivity, in-plane and out-of-plane grains misorientation as well as to probe intrinsic electrical transport properties of a high quality Bi-2223 superconductor.

Similar to other high-$T_c$ superconductors a magnetic phase diagram of Bi-2223 is roughly divided in two parts: low temperature vortex solid with a nonzero $J_c$ and a broad high-temperature dissipative vortex liquid phase. The nature of the onset of resistive transition in Bi-2223 in a magnetic field has been studied in epitaxial thin films [2], c-axis oriented Bi-2223/Ag tapes [3], Bi-2223/Bi-2212 intergrowth single crystal [4], Pb-doped Bi-2223 single crystals [5] where the vortex-glass-liquid phase transition has been observed. However, study of the vortex-liquid properties in a high quality Bi-2223 crystal has been restricted due to a lack of proper samples.

Here we report on measurements of electrical transport properties of the high quality Bi-2223 crystal below $T_c$. Bi-2223 is well known as a compound that is extremely difficult to synthesize in a single phase due to formation of intergrowth of the multiple phases. To grow high quality Bi-2223 crystals we used a modified KCl flux technique [6]. Crystals growth was performed from a nominal powder composition $Bi_{2.5}Sr_2Ca_2Cu_3O_{10+x}$ at a temperature of about 870°C in magnesia crucibles. After growth, the KCl flux was washed out by soaking the crucibles in distilled water. Collected from the crucible crystals were further annealed at 400°C in air for 20

h. Crystals of the typical size of 200x50x1 $\mu m^3$ grown in different batches have been chosen for measurements. Four linear arranged electrical contacts were prepared by silver paste with a heat treatment for about 10 min under the same conditions as during the post-annealing process, giving room temperature contact resistances below 0.1 $\Omega$. To measure current-voltage response, we used usual dc-technique including a picovoltmeter.

In Fig. 1 we present zero-field superconducting resistive transitions of a few samples measured with current 100$\mu$A. For samples #1 - #3 one can see well-defined two-step transition with sharp resistance drops at about 110K and 80K clearly indicating presence of Bi-2223 and Bi-2212 phases in these samples. Sample #4 demonstrates the one step transition where resistance falls below an instrument resolution limit at about 105K. For this sample $T_c$ defined as a mid-point of the transition is around 108K with the transition width (10-90%) nearly 2K. For further measurements we used only crystals like sample #4 where we observe single step resistive transition.

Fig. 2 shows resistance as a function of temperature for Bi-2223 crystal in fields up to 6T applied perpendicular to the CuO planes. Increase of magnetic field results in a substantial broadening of the superconducting transition similar to recently reported data for Pb-doped Bi-2223 single crystals [5] where the vortex-glass-liquid phase transition [7] was clearly identified. According to the vortex-glass model [7] in a critical region, resistance should follow power law dependence on temperature $R \propto (T-T_g)^{\nu(z-1)}$, where $\nu$ and $z$ are static and dynamic critical exponents correspondingly, $T_g$ is the vortex-glass melting transition temperature. It is natural to suggest the same vortex liquid behavior just above resistance onset with the power law resistance dependence on temperature in the high quality Bi-2223 crystal. Fits to this expression with fitting parameters $T_g$, $\nu(z-1)$, and $R_0$ are also shown in Fig.2. Table summaries data for $T_g$ and $\nu(z-1)$ at several fields. Similar to Pb-doped Bi-2223 single crystals [5] $\nu(z-1)$ within experimental error is field independent while $T_g$ monotonously decreases with increasing field.

In Fig.3 we construct a magnetic phase diagram of the high quality Bi-2223 crystal. For a comparison previously reported data for Pb-doped Bi-2223 single crystal [5] are also plotted. The vortex-glass-liquid transition line for pure Bi-2223 crystal lies on the magnetic phase diagram below similar line for Pb-doped Bi-2223 single crystal. This observation looks reasonable since concentration of pinning centers should be higher in (Bi,Pb)-2223 single crystal due to Pb doping, thus, resulting in an increase of the vortex-glass melting temperature. Also, this result is in an agreement with the previous report of substantial increase of irreversibility field in heavily Pb-doped Bi-2212 single crystal [8].

Very small cross-sectional area of our Bi-2223 crystals of about of $10^{-10}$-$10^{-11}$ m$^2$ provides an unique opportunity to probe the critical current density down to sufficiently low temperatures around 40K using currents below 0.1A. In our experiment the IV-curves were measured point by point. At any point up to the maximal current value of about 90mA used in our measurements at T=40K and B=0.02T we did not observe an increase of a voltage response with time during a few minutes at a fixed current magnitude. Also, all the IV-curves were completely reversible. These two results allow us to feel certain of absence of overheating of samples during measurements.

As an example in Fig.4 we present a set of the IV-curves obtained at T=40K at various magnetic fields. The IV-curves demonstrate well-defined negative curvature at low fields while their curvature smoothly changes from negative to positive with field increase. A straight line at field of about 3.5T corresponds to the vortex-liquid-to-solid transition, and according to the vortex-glass model [7] at higher fields the critical current density is exactly zero. Fig. 5 shows field dependence of $J_c$ defined with a voltage criterion $V=10^{-8}V$ at several temperatures [9]. Surprisingly, $J_c$ values obtained in our study for high quality Bi-2223 crystals are close to $J_c$ values reported for Bi-2223/Ag conductors [10-12]. Usually single-crystalline materials show reduced critical current density due to low density of pinning centers. On the other hand, in single crystals effect of such extrinsic properties as grains connectivity and grains misorientation on the critical current density is absent. Thus, taking into account the same order of $J_c$ magnitude in our single-crystalline like Bi-2223 samples and Bi-2223 conductors with the highest reported values of $J_c$ [10-12], we can suppose that in both cases the intrinsic low pinning strength and high anisotropy of Bi-2223 compound [13] are the main factors limiting $J_c$ while in high quality Bi-2223/Ag tapes effect of extrinsic properties on $J_c$ is already reduced. Here we should remark that Bi-2223 tapes are usually Pb-doped. Therefore it's more correctly to compare critical current density of Bi-2223 tapes with $J_c$ values of (Bi,Pb)-2223 single crystals. Our data for the vortex-glass transition line for $Bi_2Sr_2Ca_2Cu_3O_{10+x}$ and $Bi_{1.7}Pb_{0.3}Sr_2Ca_2Cu_3O_{10+x}$ crystals shown in Fig. 3 demonstrate slightly increased pinning strength in (Bi,Pb)-2223 suggesting some increase of $J_c$ values in Pb-doped crystals. Thus, we cannot exclude a possibility of some influence of extrinsic effects in Pb-doped Bi-2223 tapes. However, concluding this discussion, we note that further increase of the critical current density in Bi-2223 conductors in the first instance claims for an enhancement of amount of efficient pinning centers in this material using e.g. irradiation-induced columnar defects, heavy Pb-doping or some other variations of chemical composition of Bi-2223 compound.

In summary, we have measured the current-voltage response of the high quality $Bi_2Sr_2Ca_2Cu_3O_{10+x}$ crystal below $T_c$ in magnetic fields B//c-axis up to 6T. Similar to (Bi,Pb)-2223 single crystal the vortex liquid behavior in Bi-2223 crystal finds satisfactory description within the vortex-glass model. The vortex-glass-liquid transition line for undoped Bi-2223 crystal obtained from the vortex-glass analysis shifted to lower temperature compared to Pb-doped Bi-2223 single crystal reflecting reduced pinning strength. The fact that the critical current density of high quality Bi-2223 crystals is of the same order of magnitude as $J_c$ values of the best samples of Bi-2223 tapes, allows us to consider weak intrinsic pinning properties of Bi-2223 as a main limiting factor of $J_c$ in both Bi-2223 crystals and Bi-2223 conductors.

Acknowledgements

This work is supported by the New Energy and Industrial Technology Development Organization (NEDO) as Collaborative Research and Development of Fundamental Technologies for Superconductivity Applications and partially supported by the Russian Federal Agency on science and innovation (02.513.11.3378).

**Legends to Figures:**

**Fig.1**

Zero-field temperature dependence of normalized resistance for several Bi-2223 crystals with different level of Bi-2212 intergrowth. Within experimental resolution there are no signs of Bi-2212 intergrowth in sample #4.

**Fig.2**

Temperature dependence of resistance at various magnetic fields. Lines show fits to vortex-glass transition model with fitting parameters listed in Table.

**Fig.3**

Vortex-glass transition line for high quality Bi-2223 crystal. For comparison data from Ref.5 for Pb-doped Bi-2223 single crystals are also shown. Dashed line is guide for the eye.

**Fig.4**

IV-curves at T=40K and B=0.02T, 0.05T, 0.1T, 0.2T, 0.5T, 0.7T, 1T, 1.3T, 1.5T, 1.7T, 2T, 2.5T, 3T, 3.5T, 4T, 4.5T, 5T, 5.5T. Increasing field magnitudes correspond to curves from right to left. Dashed lines are guides for the eye. Solid line corresponds to vortex-liquid-to-solid-transition at about 3.5T.

**Fig.5**

Field dependence of the critical current density at various temperatures. Solid lines are guides for the eye.

**Table**

Vortex-glass-liquid phase transition temperature $T_g$ and critical exponent $\nu(z-1)$ at several fields.

| Magnetic field (T) | $T_g$ (K) | $\nu(z-1)$ |
|:---:|:---:|:---:|
| 0.1 | 78.2±5.0 | 5.3±2.4 |
| 0.3 | 66.5±3.0 | 5.1±1.2 |
| 1.0 | 49.3±3.0 | 4.9±1.0 |
| 3.0 | 38.2±2.0 | 5.0±0.4 |
| 6.0 | 34.3±2.0 | 4.6±0.4 |

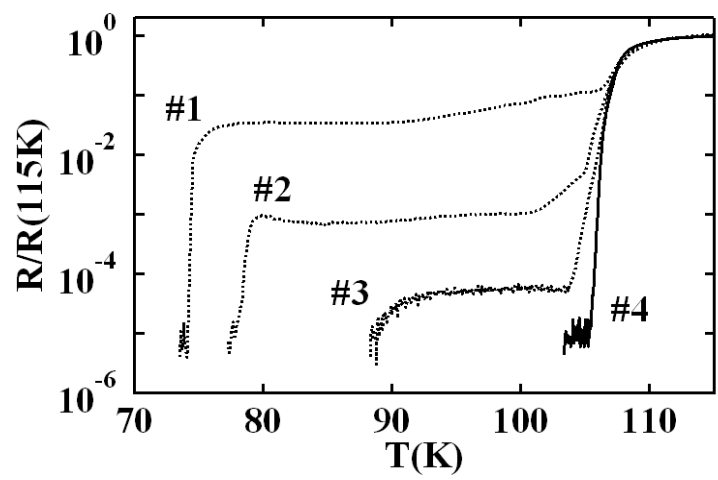

**Fig.1**

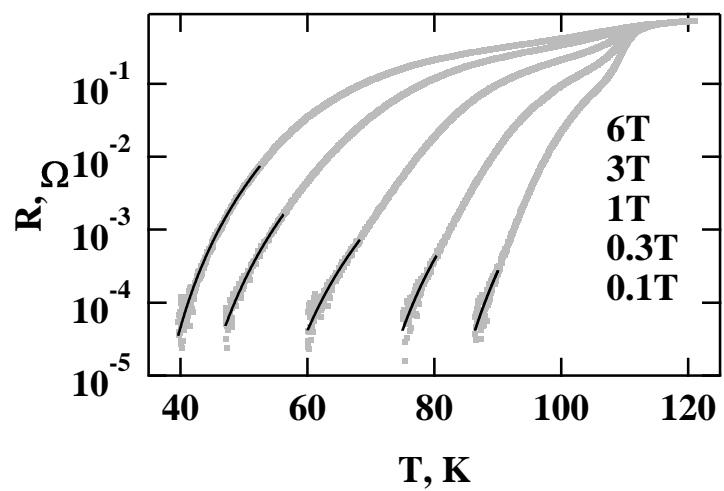

**Fig.2**

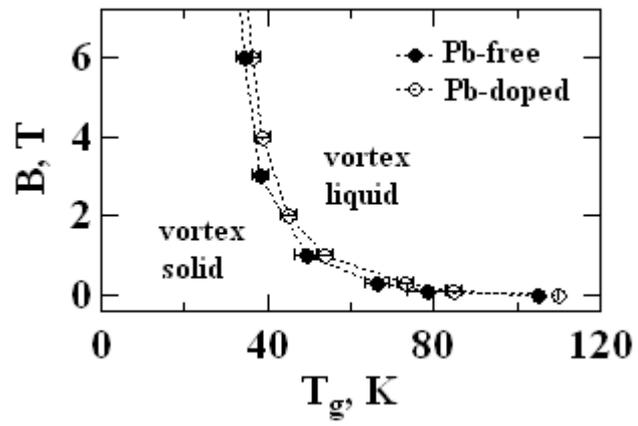

**Fig.3**

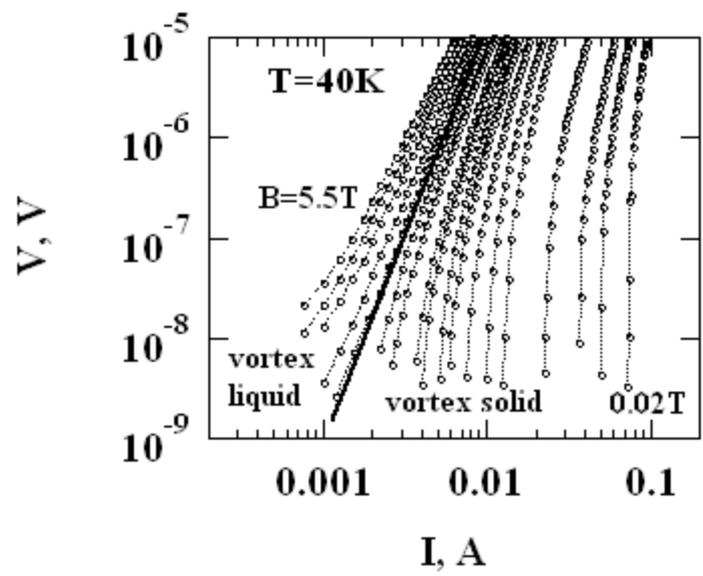

**Fig.4**

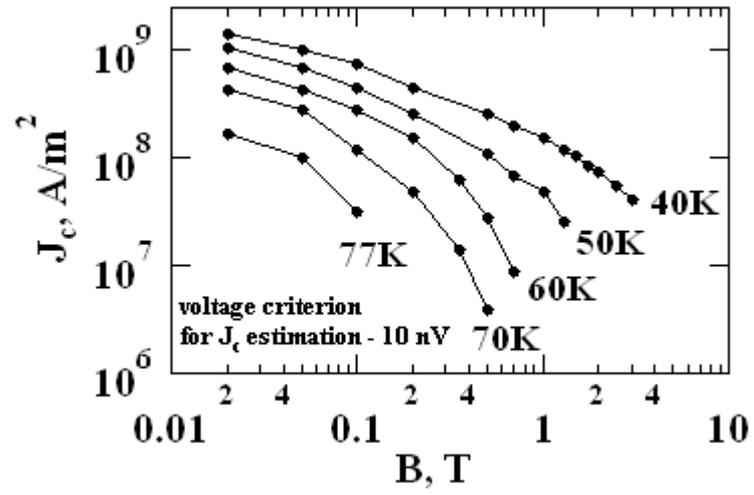

**Fig.5**